# Necessary and Sufficient Elastic Stability Conditions in Various Crystal Systems

Félix Mouhat and François-Xavier Coudert*

*PSL Research University, Chimie ParisTech – CNRS,  
Institut de Recherche de Chimie Paris, 75005 Paris, France*  
(Dated: December 5, 2014)

While the Born elastic stability criteria are well-known for cubic crystals, there is some confusion in the literature about the form it should take for lower symmetry crystal classes. We present here closed form necessary and sufficient conditions for elastic stability in all crystal classes, as a concise and pedagogical reference to stability criteria in non-cubic materials.

## I. INTRODUCTION

The fundamental understanding of the conditions of mechanical stability of unstressed crystalline structures dates back to the seminal work of Max Born and co-authors in the 1940s,[1] and was consolidated in his 1954 book.[2] This and later textbooks[3–5] usually state the generic requirements for elastic stability of crystal lattices, and give simplified equivalents of the generic conditions for some high-symmetry crystal classes. In particular, in the case of cubic crystals, the conditions of stability reduce to a very simple form:

$$C_{11} - C_{12} > 0 \; ; \; C_{11} + 2C_{12} > 0 \; ; \; C_{44} > 0 \qquad (1)$$

The above equations for the cubic crystal system are well-known, and often called the "Born stability criteria". We noticed however, through a review of the recent literature on the experimental measurements and first principles calculations of elastic constants of solids, that there is a large amount of confusion about the form that these conditions should take for other crystal classes, including hexagonal, tetragonal, rhombohedral and orthorhombic classes. In more than a few cases, incorrect generalizations of the cubic criteria have been published;[6–10] this is particularly frequent for orthorhombic crystals.[11–17] In other papers, the authors rely on conditions that are necessary but not sufficient.[18] So long as the diagonal elastic constants $C_{ii}$ are dominant, this leads to wrong quantitative analyses, but does not change the qualitative picture (whether a specific crystal is stable or not). However, we identified at least one case where accounting for the proper stability criteria did change the conclusions drastically, meaning that a system (MOF-74 material with $CH_4$ guest molecules) was identified as stable when it is not.[10]

In this short paper, we summarize the generic elastic stability conditions for crystals, and present necessary and sufficient conditions for each crystal class. We also detail the crystal classes where no analytical necessary and sufficient conditions exist.

## II. GENERAL ELASTIC STABILITY CONDITION

The elastic behavior of a lattice are described by its matrix of second-order elastic constants:

$$C_{ij} = \frac{1}{V_0} \left( \frac{\partial^2 E}{\partial \varepsilon_i \partial \varepsilon_j} \right) \qquad (2)$$

where $E$ is the energy of the crystal, $V_0$ its equilibrium volume and $\varepsilon$ denotes a strain.[19] This elastic matrix (also called stiffness matrix) has size $6 \times 6$ and is symmetric: it is thus composed of 21 independent components. The crystal class of the material considered yields additional symmetry constraints, further reducing the number of independent elastic constants. For arbitrary homogeneous deformation by an infinitesimal strain, the energy of the crystal is therefore given by the following quadratic form:

$$E = E_0 + \frac{1}{2} V_0 \sum_{i,j=1}^{6} C_{ij} \varepsilon_i \varepsilon_j + O\left(\varepsilon^3\right) \qquad (3)$$

A crystalline structure is stable, in the absence of external load and in the harmonic approximation,[20] if and only if *(i)* all its phonon modes have positive frequencies for all wave vectors (*dynamical stability*), and *(ii)* the elastic energy, given by the quadratic form of Eq. 3, is always positive ($E > 0, \forall \varepsilon \neq 0$). This latter condition is called the *elastic stability* criterion. As first noted by Born,[1] it is mathematically equivalent with the following *necessary and sufficient* stability conditions:

- the matrix **C** is definite positive;
- all eigenvalues of **C** are positive;
- all the leading principal minors of **C** (determinants of its upper-left $k$ by $k$ sub-matrix, $1 \leq k \leq 6$) are positive, a property known as Sylvester's criterion;
- an arbitrary set of minors of **C** are all positive. It can be useful to choose, for example, the *trailing* minors, or any other set.

These are four possible formulations of the generic *Born elastic stability conditions* for an unstressed crystal. They are valid regardless of the symmetry of the crystal studied, and are not linear.

Finally, from these conditions we can deduce some *necessary but not sufficient* conditions. Fedorov,[4] in particular, noted that the condition on principal minors implies that all diagonal elements are positive ($C_{ii} > 0, \forall i$), but this alone is not strong enough to ensure stability. Another example of *necessary* condition is

$$(C_{ij})^2 < C_{ii} C_{jj} \quad \forall i, j \qquad (4)$$

### III. EXPRESSIONS FOR SPECIFIC LAUE CLASSES

We now turn our attention to express closed form expressions of the *necessary and sufficient* elastic stability conditions for 11 Laue classes, as described in Table I. For each class, the Table also gives the number of *independent* elastic constants in the stiffness matrix. We focus here on the crystalline systems in three dimensions, but this analysis can also be extended in very similar terms to other dimensions (e.g., one- and two-dimensional quasicrystals).[21,22]

For each crystal system and Laue class, closed form expressions of the necessary and sufficient elastic stability conditions can be found following a number of different approaches. The one we have chosen is to develop the series of minors of the stiffness matrix, in an order chosen to minimize the degree of the polynomials involved. For this, one reorders the matrix into block diagonal form and expresses the minors starting with the smallest blocks. Another way to view this approach is to express the quadratic form of the energy and reducing it by successively "completing the square" in the variables, taken in a sequence appropriate to the symmetries.[23,24] This is formally equivalent and gives identicals expressions for the conditions. Finally, we also checked the results presented below by means of direct calculation with computer algebra sofware,[25] expanding the characteristic polynomial of the stiffness matrix in each case, and factoring it.

#### A. Cubic crystal system

The cubic crystal system has the simplest form of elastic matrix, with only 3 independent constants: $C_{11}$, $C_{12}$ and $C_{44}$:

$$\boldsymbol{C}_{\text{cubic}} = \begin{pmatrix} C_{11} & C_{12} & C_{12} & & & \\ . & C_{11} & C_{12} & & & \\ . & . & C_{11} & & & \\ & & & C_{44} & & \\ & & & & C_{44} & \\ & & & & & C_{44} \end{pmatrix} \quad (5)$$

(in this notation, dots are used to indicate nonzero elements constrained by the symmetric nature of the matrix). The three Born stability criteria for the cubic system are well-known:

$$C_{11} - C_{12} > 0 \; ; \; C_{11} + 2C_{12} > 0 \; ; \; C_{44} > 0 \quad (6)$$

They are necessary and sufficient. Here we merely note that the first two conditions imply that $C_{11} > 0$, so it needs not be noted as an extra condition, as is sometimes done. Also, the first condition can be equivalently stated as $C_{11} > |C_{12}|$.

#### B. Hexagonal and tetragonal classes

Both Laue classes of the hexagonal crystal system, as well as the tetragonal (I) class ($4/mmm$), have the same form for the elastic matrix:

| Crystal system | Laue class | Point groups | $C_{ij}$'s |
|---|---|---|---|
| Triclinic | $\bar{1}$ | $1, \bar{1}$ | 21 |
| Monoclinic | $2/m$ | $2, m, 2/m$ | 13 |
| Orthorhombic | $mmm$ | $222, 2mm, mmm$ | 9 |
| Tetragonal (II) | $4/m$ | $4, \bar{4}, 4/m$ | 7 |
| Tetragonal (I) | $4/mmm$ | $4mm, 422, \bar{4}2m, 4/mmm$ | 6 |
| Rhombohedral (II) | $\bar{3}$ | $3, \bar{3}$ | 7 |
| Rhombohedral (I) | $\bar{3}m$ | $32, 3m, \bar{3}m$ | 6 |
| Hexagonal (II) | $6/m$ | $6, \bar{6}, 6/m$ | 5 |
| Hexagonal (I) | $6/mmm$ | $6mm, 622, \bar{6}2m, 6/mmm$ | 5 |
| Cubic (II) | $m\bar{3}$ | $23, m\bar{3}$ | 3 |
| Cubic (I) | $m\bar{3}m$ | $432, \bar{4}3m, m\bar{3}m$ | 3 |

Table I. Laue groups and number of independent second-order elastic constants $C_{ij}$. We follow the naming convention of Wallace[5] (I/II) to distinguish Laue classes within the same crystal system.

$$\boldsymbol{C}_{\text{hexa/tetra I}} = \begin{pmatrix} C_{11} & C_{12} & C_{13} & & & \\ . & C_{11} & C_{13} & & & \\ . & . & C_{33} & & & \\ & & & C_{44} & & \\ & & & & C_{44} & \\ & & & & & C_{66} \end{pmatrix} \quad (7)$$

Crystals of the tetragonal (I) class thus have 6 independent elastic constants, while those with hexagonal crystal system have only 5, due to the added relation:

$$C_{66} = (C_{11} - C_{12})/2 \quad (8)$$

By direct calculation of the eigenvalues of the stiffness matrix above, one can derive the following four *necessary and sufficient* conditions for elastic stability in the hexagonal and tetragonal (I) case:

$$\begin{cases} C_{11} > |C_{12}| \; ; \; 2C_{13}^2 < C_{33}(C_{11} + C_{12}) \\ C_{44} > 0 \; ; \; C_{66} > 0 \end{cases} \quad (9)$$

(where the condition on $C_{66}$ is redundant with the first one, for the hexagonal case).

The tetragonal (II) class ($4/m$) features an extra elastic constant, $C_{16}$, bringing the total of independent $C_{ij}$'s to 7:

$$\boldsymbol{C}_{\text{tetra II}} = \begin{pmatrix} C_{11} & C_{12} & C_{13} & & & C_{16} \\ . & C_{11} & C_{13} & & & -C_{16} \\ . & . & C_{33} & & & \\ & & & C_{44} & & \\ & & & & C_{44} & \\ . & . & & & & C_{66} \end{pmatrix} \quad (10)$$

The conditions for this are the same as for the tetragonal (I) class, Eq. 9, with the exception of the condition on $C_{66}$ being replaced by: $2C_{16}^2 < C_{66}(C_{11} - C_{12})$. Thus the complete *necessary and sufficient* Born stability criteria for tetragonal (II) class are:

$$\begin{cases} C_{11} > |C_{12}| \,;\, 2C_{13}^2 < C_{33}(C_{11} + C_{12}) \\ C_{44} > 0 \,;\, 2C_{16}^2 < C_{66}(C_{11} - C_{12}) \end{cases} \quad (11)$$

### C. Rhombohedral classes

Crystals in the rhombohedral (I) class (Laue class $\bar{3}m$) feature 6 independent elastic constants:

$$\boldsymbol{C}_{\text{rhombo I}} = \begin{pmatrix} C_{11} & C_{12} & C_{13} & C_{14} & & \\ . & C_{11} & C_{13} & -C_{14} & & \\ . & . & C_{33} & & & \\ . & . & & C_{44} & & \\ . & . & & & C_{44} & C_{14} \\ . & . & & & . & C_{66} \end{pmatrix} \quad (12)$$

where, like in the hexagonal case, $C_{66} = (C_{11} - C_{12})/2$. We therefore obtain the following four *necessary and sufficient* conditions:

$$\begin{cases} C_{11} > |C_{12}| \,;\, C_{44} > 0 \\ C_{13}^2 < \frac{1}{2}C_{33}(C_{11} + C_{12}) \\ C_{14}^2 < \frac{1}{2}C_{44}(C_{11} - C_{12}) = C_{44}C_{66} \end{cases} \quad (13)$$

For the rhombohedral (II) class, there is one more independent elastic constant, namely $C_{15}$:

$$\boldsymbol{C}_{\text{rhombo II}} = \begin{pmatrix} C_{11} & C_{12} & C_{13} & C_{14} & C_{15} & \\ . & C_{11} & C_{13} & -C_{14} & -C_{15} & \\ . & . & C_{33} & & & \\ . & . & & C_{44} & & -C_{15} \\ . & . & & & C_{44} & C_{14} \\ . & . & & & . & C_{66} \end{pmatrix} \quad (14)$$

The resolution leads to a similar case as the rhombohedral (I) class, but with a stricter version of the last criterion:

$$\begin{cases} C_{11} > |C_{12}| \,;\, C_{44} > 0 \\ C_{13}^2 < \frac{1}{2}C_{33}(C_{11} + C_{12}) \\ C_{14}^2 + C_{15}^2 < \frac{1}{2}C_{44}(C_{11} - C_{12}) = C_{44}C_{66} \end{cases} \quad (15)$$

The criteria presented in Eq. 15, though they are rigorously *necessary and sufficient*, have not been presented so far in the scientific literature or textbooks, to our knowledge.

### D. Orthorhombic Systems

Finally, we come to the crystal systems with lower symmetry and larger number of independent elastic constants. The stiffness matrix for an orthorhombic crystal has the following form, with 9 constants and no relationships between them:

$$\boldsymbol{C}_{\text{ortho}} = \begin{pmatrix} C_{11} & C_{12} & C_{13} & & & \\ . & C_{22} & C_{23} & & & \\ . & . & C_{33} & & & \\ & & & C_{44} & & \\ & & & & C_{55} & \\ & & & & & C_{66} \end{pmatrix} \quad (16)$$

There are three trivial eigenvalues for this matrix, namely $C_{44}$, $C_{55}$ and $C_{66}$, all of which need to be positive. However, the eigenvalues of the upper-left $3 \times 3$ block do not have closed form expression. They are the three roots of the following cubic polynomial:

$$\begin{aligned} \lambda^3 &- \lambda^2 \left( C_{11} + C_{22} + C_{33} \right) \\ &+ \lambda \left( C_{11}C_{22} + C_{11}C_{33} + C_{22}C_{33} - C_{12}^2 - C_{23}^2 - C_{13}^2 \right) \\ &+ C_{11}C_{23}^2 + C_{22}C_{13}^2 + C_{33}C_{12}^2 \\ &- C_{11}C_{22}C_{33} - 2C_{12}C_{13}C_{23} \end{aligned} \quad (17)$$

One can, however, find a closed form expression equivalent to the generic criteria, through the requirement that leading principal minors be positive. This reduces to the following *necessary and sufficient* Born criteria for an orthomrhombic system:

$$\begin{cases} C_{11} > 0 \,;\, C_{11}C_{22} > C_{12}^2 \\ C_{11}C_{22}C_{33} + 2C_{12}C_{13}C_{23} \\ \quad - C_{11}C_{23}^2 - C_{22}C_{13}^2 - C_{33}C_{12}^2 > 0 \\ C_{44} > 0 \,;\, C_{55} > 0 \,;\, C_{66} > 0 \end{cases} \quad (18)$$

The conditions obtained are not all linear, but polynomial functions of the elastic constants (because the largest non-diagonal block in the stiffness matrix has size $3 \times 3$ and all coefficients are independent).

We feel it is important to note here that some authors have presented in the literature simpler stability conditions for orthorhombic crystals, many of them linear![6,11,12] To quote only one, Wu et al.[6] claim that "it is known that for orthorhombic crystals, the mechanical stability requires the elastic stiffness constants satisfying the following conditions":

$$\begin{cases} C_{ii} > 0 \,;\, C_{ii} + C_{jj} - 2C_{ij} > 0 \\ C_{11} + C_{22} + C_{33} + 2(C_{12} + C_{13} + C_{23}) > 0 \end{cases} \quad (19)$$

These conditions seem like a natural extension of the well-known cubic case, but they are incorrect. Indeed, it is easy to verify formally, with CAS (Computer Algebra System) software, that these conditions are *necessary but not sufficient*.[26]

### E. Monoclinic and Triclinic Systems

Monoclinic and triclinic crystal systems have 13 and 21 independent elastic constants respectively. Given the complexity of the equations obtained, we will not show those here. When studying such low-symmetry crystals, it is usually more convenient to keep the stiffness coefficents in matrix form. In particular, the generic *necessary and sufficient* criterion that all eigenvalues of $C$ be positive is easily to check with simple linear algebra routines.

If, nevertheless, one wishes to obtain closed form expressions for the stability conditions of monoclinic and triclinic systems, they can be obtained as 6 polynomials in the elastic constants by writing out that the leading principal minors of $C$ be positive. For monoclinic systems, the polynomials will be of degree 4 (at most), while for triclinic crystals they will be of degree 6 (at most). Simpler forms, including fully linear or quadratic, that have sometimes proposed in the literature,[6] are incorrect.

## IV. CONCLUSION

We gathered here, for the first time, closed form expressions for necessary and sufficient elastic stability criteria (also called Born stability conditions) depending on the Laue classes of crystals. While high symmetry crystal systems allow for simple formulas, these cannot be generalized trivially to lower symmetry systems. In particular, the cubic system is the only one for which these conditions are all linear. Hexagonal, tetragonal and rhombohedral systems have quadratic stability criteria. Conditions for orthorhombic crystals involve cubic polynomials, while monoclinic and triclinic systems can be expressed as quartic and sextic polynomials, respectively.

Finally, we note that the conditions of elastic stability described here for an unstressed system can be readily generalized to systems under an arbitrary external load $\sigma$ by introducing an elastic stiffness tensor $B$ under load, defined as (in tensorial notation):[20,27]

$$B_{ijkl} = C_{ijkl} + \frac{1}{2}\left(\delta_{ik}\sigma_{jl} + \delta_{jk}\sigma_{il} + \delta_{il}\sigma_{jk} + \delta_{jl}\sigma_{ik} - 2\delta_{kl}\sigma_{ij}\right) \quad (20)$$

The resulting symmetry of the $B$ tensor might be lower than that of $C$, if the load is not isotropic. Elastic stability conditions can then be derived, as a function of crystal system and symmetry of the external load, by applying the formalism of this paper.[28]

## ACKNOWLEDGMENTS

We thank Dr. David Bousquet for cokeful discussions.